# Topological insulator $Bi_2Se_3$ thin films as an alternative channel material in MOSFETs

Jiwon Chang, Leonard F. Register and Sanjay K. Banerjee

Microelectronics Research Center, The University of Texas at Austin, Austin, TX 78758, USA

Three-dimensional (3-D) topological insulators (TI) are characterized by the presence of metallic surface states and a bulk band gap. Recently theoretical and experimental studies have shown an induced gap in the surface state bands of TI thin films. The gap results from interaction of conduction band (CB) and valence band (VB) surface states from the opposite surfaces of a thin film, and its size is determined by the film thickness. This gap formation could open the possibility of thin-film TI-based metal-oxide-semiconductor field-effect transistors (MOSFETs). Here we explore the performance of MOSFETs based on TI thin films, specifically $Bi_2Se_3$, using quantum ballistic transport simulations with the tight-binding Hamiltonian in the atomic orbital basis. Our simulations indicate that $Bi_2Se_3$ MOSFET will be vulnerable to short-channel effects due to the high relative dielectric constant of $Bi_2Se_3$ (~100) despite its expected excellent electrostatic integrity inherent in a two-dimensional system, and will have other limitations as compared to silicon–based MOSFETs. However, $Bi_2Se_3$ MOSFETs, and presumably other TI-based MOSFETs, appear to provide reasonable performance that perhaps could provide novel device opportunities when combined with novel TI properties such as spin-polarized surface states.

# I.  INTRODUCTION

Three-dimensional (3-D) topological insulators (TIs) have attracted great attention recently because of their novel electronic surface states [1,2,3]. A 3-D TI is characterized by the presence of protected spin-polarized semi-metallic surface states with the conduction band (CB) and valence band (VB) meeting at a Dirac point, separated by an insulating bulk. Conducting surface states of TI are quite robust to nonmagnetic disorder but open a gap in the presence of time-reversal symmetry breaking perturbations [4,5,6]. However, recent theoretical and experimental studies have shown an induced gap within the surface bands in TI thin films even without magnetic disorder [7,8,9,10,11]. The gap originates from allowed VB-to-CB surface states interactions between the opposite surfaces, with a gap size determined by the thin film thickness. (CB-to-CB and VB-to-VB inter-surface interactions are essentially forbidden by their orthogonal spin status.) This gap opening allows for the possibility of TI-based metal-oxide-semiconductor field-effect transistors (MOSFETs). Experiments have demonstrated FETs using a TI thin film and shown a gate-tunable conductance [12,13,14,15,16,17]. There have been several theoretical studies of transport in TIs using a conceptual model Hamiltonian for the metallic surface bands [18,19,20]. However, for a more accurate treatment including the effects of gap opening, we use a full-band treatment of the TI thin-film band structure. Specifically, in this work, we use quantum ballistic non-equilibrium Green's function (NEGF) simulations with the atomic orbital-based tight-binding (TB) Hamiltonian obtained from density functional theory (DFT) to explore the performance of MOFET based on a thin film of the TI $Bi_2Se_3$.

# II.  COMPUTATIONAL APPROACH

Figure 1 shows a hexagonal unit cell of bulk $Bi_2Se_3$ with lattice parameters $a$ = 0.4138 nm and $c$ = 2.8633 nm. The building block of the hexagonal bulk $Bi_2Se_3$ crystal consists of five atomic layers referred to as a quintuple layer (QL). The square shaded region in Figure 1 shows one such QL. The entire structure of Figure 1 contains three QLs, i.e., 15 atomic layers stacked along the z-direction. The atomic planes within a QL are arranged in a sequence Se1-Bi-Se2-Bi-Se1 where the "1" and "2" indicate different Se layer structures. A thin film structure can be formed from one up to a stack of several QLs.

Surface band structures for 6QLs, 3QLs, 2QLs and 1QL $Bi_2Se_3$ thin films are shown in Figure 2. Band structures are obtained from DFT calculations using the OPENMX code [21], based on a linear combination of pseudoatomic orbital (PAO) method. The pseudopotentials were generated from full relativistic calculations, and the generalized gradient approximation was applied for the exchange-correlation potential [22]. Metallic surface states within a bulk band gap (~300 meV) exist in the 6 QLs thin film. However, a non-zero gap is produced in the 3 QLs, 2QLs and 1QL thin films resulting from interactions between CB and VB states from nominally opposite surfaces. The gap size increases rapidly as the thin film thickness is reduced, reaching about 497 meV in the 1QL thin film. In this work, we consider only the 1QL thin film, since it has the largest band gap, with a value that is technologically interesting for electronics.

For the transport calculation, we define a series of rectangular unit cells in the simulation region, oriented perpendicular to the transport direction $x$, as shown Figure 3(a) in the top view of a thin film. Three different symbols (○, □ and ×) are for atomic positions in the different atomic layers stacked in the z-direction. The TB hopping potentials used in the transport calculations are extracted from DFT using maximally localized Wannier functions (MLWFs) [23]. At least 3$^{rd}$ nearest neighbor coupling is used to accurately reproduce the DFT-obtained band structure. Figure 3(b) shows the simulated device structure of a 1QL $Bi_2Se_3$ MOSFET. Semi-infinite source and drain are n-type doped to $1\times10^{13}$ cm$^{-2}$, corresponding to the Fermi-level about 50 meV above the CB edge. The undoped channel is gated using a 2 nm physical thickness $HfO_2$ (dielectric constant $\kappa = 25$) gate insulator. We consider two different channel lengths of about 20 nm and 50 nm, respectively. A relative dielectric constant of 100 is used for $Bi_2Se_3$ [24]. The work function of undoped $Bi_2Se_3$ is assumed to be same as that of the gate for simplicity. Electron probability density is injected into the simulation region from the set of incident propagating eigenmodes of semi-infinite leads, subdivided by the TI-plane-normal ($z$) subbands, and in-plane incident and transverse wave-vectors, $k_x$ and $k_y$, respectively, weighted by the Fermi distribution function of the injecting lead. We use recursive scattering matrices to propagate injected probability from the source (drain), through the channel, to the drain (source), and/or to reflect it back to the source (drain) [25]. The transport calculation is performed iteratively with a Poisson solver until self-consistency between charge density and electrostatic potential is achieved. These simulations ignore the effect of various scattering mechanisms. The electron-phonon scattering on the surface of $Bi_2Se_3$ may place additional limits on device performance, and phonon-assisted band-to-band tunneling in the channel/drain junction could increase the subthreshold leakage current [26,27].

## III. RESULTS AND DISCUSSION

Simulation results for the 50 nm channel length device are shown in Figure 5. From the transfer characteristics ($I_{DS}$ vs. $V_{GS}$) in Figure 5(a), the maximum current at $V_{GS} = 0.7$ V is about 1.1 mA/μm. The ratio of maximum current (at $V_{GS} = 0.7$ V) to minimum current (at $V_{GS} = -0.6$ V) is more than $10^{12}$. The subthreshold slope is about 65 mV/dec, close to the ideal value of 60 mV/dec, and drain-induced barrier lowering (DIBL) is fairly small (~40 mV/V). Output characteristics ($I_{DS}$ vs. $V_{DS}$) at three different values of $V_{GS}$ in Figure 5(b) indicate the saturation of drain currents beyond $V_{DS}$ = 0.2 V. In Figure 5(c), $I_{ON}$ is plotted as a function of the ratio between ON current at $V_{ON}$ and OFF current at $V_{OFF}$ ($I_{ON}/I_{OFF}$) at the power supply voltage $V_{DD}$ = 0.5 V ($V_{ON} - V_{OFF} = V_{DD}$). At $I_{ON}/I_{OFF} = 10^4$, $I_{ON}$ is about 600 µA/µm. The transconductance ($g_m$ = $\partial I_{DS}/\partial V_{GS}$) variation with $V_{GS}$ at $V_{DS}$ = 0.5 V is also plotted in Figure 5(d). The transconductance monotonically increases along with $V_{GS}$ and reaches its maximum value 2.8 mS/µm around $V_{GS}$ = 0.65 V. Figure 6(a) and 6(b) show CB and VB edges profiles along the channel direction at $V_{DS}$ = 0.05 V and $V_{DS}$ = 0.5 V, for different $V_{GS}$ from −0.4 V to 0.5 V in steps of 0.1 V, respectively. From the band edge profiles, we can observe that the potential changes very slowly along the device due to the extremely high dielectric constant of $Bi_2Se_3$ (~100). Potential variations between the channel and source/drain occur over a long distance, such that there is no flat potential profile in the channel. The potential profile in the entire channel region is rounded, which suggests that the lateral electric field penetrates through the channel from drain to source. This deep penetration of lateral field into the channel is problematic in a short channel device, the details of which will be discussed later for a 20 nm channel length MOSFET. However, in for this channel length, the high dielectric constant might be helpful to suppress gate-induced drain leakage (GIDL), one of the main leakage mechanisms limiting low off-state current. In Figure 6(b), even if there exists an overlap between CB and VB in the region between the channel and drain with $V_{GS}$ in the range from -0.1 to -0.4 V, band-to-band tunneling is negligible because of the thick tunnel barrier.

A 20 nm channel length device of 1QL $Bi_2Se_3$ thin film was simulated otherwise using the same parameters as a 50 nm channel length device. The key device characteristics, $I_{DS}$ vs. $V_{GS}$ and $I_{DS}$ vs. $V_{DS}$, are shown in Figure 7(a) and 7(b). The overall device performance is significantly degraded compared with the 50 nm device. There exists a substantial amount of current flow in the subthreshold regime. The minimum current at $V_{GS} = -0.6$ V increases by a factor of $10^5$ and the subthreshold slope is about 110 mV/dec. Short channel effects also become likely intolerable in the 20 nm device, with severe DIBL (~330 mV/V) and threshold voltage $V_T$ roll-off (Figure 7(a)). Poor saturation is seen in the $I_{DS}$ vs. $V_{DS}$ curves of Figure 7(b). The $I_{ON}/I_{OFF}$ vs. $I_{ON}$ characteristic at $V_{DD} = 0.5$ V of Figure 7(c) exhibits poor ON-OFF ratios. The achievable $I_{ON}$ at $I_{ON}/I_{OFF} = 10^4$ is about 260 µA/µm, lower than that of a 50 nm channel length device in Figure 5(c). Poor subthreshold behavior and severe short channel effects can be explained by examining the band edge profiles shown in Figure 8(a) and 8(b). From Figure 8(a), the maximum of CB edge at $V_{GS} = -0.4$ V along the channel direction is reduced to about 275 meV, as compared to about 525 meV for the 50 nm channel case (Figure 6(a)). The maximum of CB edge for $V_{GS} = -0.4$ V is pulled down substantially by increasing $V_{DS}$ from $V_{DS} = 0.05$ V (Figure 8(a)) to $V_{DS} = 0.5$ V (Figure 8(b)), indicating substantial DIBL. The reason for these poor characteristics is the relatively slow variation of the potential along due to the extremely high dielectric constant of $Bi_2Se_3$ (~100), even though the TI layer itself is thin. The use of the high-k dielectric, intended to provide better gate control in modern devices, may also be of mixed value.

It is instructive to compare key characteristics of 1QL $Bi_2Se_3$ MOSFET with those of conventional Si MOSFETs. We compare the 1QL $Bi_2Se_3$ MOSFET with the silicon-on-insulator (SOI) MOSFET with comparable device parameters. Ballistic quantum transport simulations of the SOI MOSFET are performed with the real-space effective mass Hamiltonian [28]. Device parameters of the simulated SOI MOSFET are taken from a previous experimental study [29], and are as follows: channel length = 30 nm, effective oxide thickness = 1.3 nm and n-type doping density of source and drain = $5 \times 10^{12}$ cm$^{-2}$. The same parameters are then used in the simulation of 1QL $Bi_2Se_3$ MOSFET. Figure 9 shows the comparison of key device performance parameters of the two simulated devices. The overall performance is better in the SOI MOSFET. As shown from transfer characteristics ($I_{DS}$ vs. $V_{GS}$) in Figure 9(a), the SOI MOSFET is superior to the 1QL $Bi_2Se_3$ MOSFET in terms of the subthreshold slope (60 meV/dec and 130 meV/dec for SOI and $Bi_2Se_3$ MOSFETs, respectively). For the same gate stack, a larger gate bias is required to induce the same potential shift in the channel. DIBL is also more severe in the $Bi_2Se_3$ MOSFET (130 meV/V and 180 meV/V for SOI and $Bi_2Se_3$ MOSFETs, respectively). The lateral electric field reaches further in $Bi_2Se_3$ than in Si, in turn, lowering the channel potential more. Large subthreshold slope and DIBL result in the poor ON-OFF ratio in the 1QL $Bi_2Se_3$ MOSFET (Figure 9(b)). For $I_{ON}/I_{OFF} = 10^4$, almost 100 times larger ON current is achievable in the SOI MOSFET. Over the save gate voltage range above threshold, the maximum obtainable transconductance before current becomes lead-limited is more than 8 mS/μm in the SOI MOSFET but only 1.2 mS/μm in the $Bi_2Se_3$ MOSFET. The quantum (channel density of states) capacitance in the $Bi_2Se_3$, which has only one occupied "subband" within its one energy valley—much like for thin-channel III-V MOSFETs but with a higher conductivity effective mass—is substantially less than for multi-valley Si, while the conductivity effective mass is comparable, so that the transconductance should be somewhat less even absent short-channel effects.

## IV. CONCLUSION

In summary and conclusion, we performed quantum ballistic transport simulations using a tight-binding Hamiltonian in the atomic orbital basis to assess the feasibility of $Bi_2Se_3$ for an alternative channel material in MOSFETs. A 1QL (~0.7 nm) $Bi_2Se_3$ thin film with the largest band gap (~497 meV) is considered. We first investigated two different channel length devices, 20 nm and 50 nm. The 1QL $Bi_2Se_3$ MOSFET with 50 nm channel length exhibits good subthreshold device performance with a subthreshold slope close to 60 meV, as well as small

DIBL. In the 20 nm channel device, however, these device characteristics for the $Bi_2Se_3$ MOSFET are downgraded severely. The reason for the poor performance is the very high dielectric constant of $Bi_2Se_3$. The large dielectric constant slows the variation of the potential along the device making the MOSFET more vulnerable to short channel effects. However, even the on-state performance of long-channel $Bi_2Se_3$ MOSFETs is limited. As seen the 30 nm $Bi_2Se_3$ MOSFET performs poorly in terms of transconductance as compared to the 30 nm Si MOSFET in a head-to-head comparison (Figure 9(c)). Moreover, so does the 50 nm—again, ballistic—$Bi_2Se_3$ MOSFET by comparison to the Si MOSFET over the same approximately 0.4 V $V_G - V_{th}$ range before lead limited current flow takes over in the $Bi_2Se_3$ MOSFET, despite reduced gate capacitance for the Si device.

Of course, the results here are specific to $Bi_2Se_3$ MOSFETs. Its high dielectric constant clearly illustrates qualitative short-channel electrostatic effects of using even thin films of high-k TIs. However, other TIs with smaller dielectric constants should fare quantitatively better in terms of short channel effects. $TlBiSe_2$, for example, has an at least smaller dielectric constant of 25 [30]. That said (as we also obtained from DFT calculations as described above) $TlBiSe_2$, itself, has both a higher conductivity effective mass (~0.7), suggesting poorer on-state performance devices, and a smaller band gap (~300 meV), suggesting the possibility of increased off-state leakage, than $Bi_2Se_3$ for comparable (~1.0 nm) thicknesses [31]. Still, there may be better TI candidates with a better combination of properties.

On the other hand, this work suggests that $Bi_2Se_3$ MOSFET, and likely other TI-based MOSFETs, could still provide reasonable performance, even if not state-of-the art silicon-MOSFET-like. That performance combined with the novel properties of TI such as spin-polarized surface states perhaps could provide novel opportunities that Si or other conventional semiconductors cannot.

ACKNOWLEDGMENT

The authors acknowledge support from the Nanoelectronics Research Initiative supported Southwest Academy of Nanoelectronics (NRI-SWAN) center. We thank the Texas advanced computing center (TACC) for computational support.

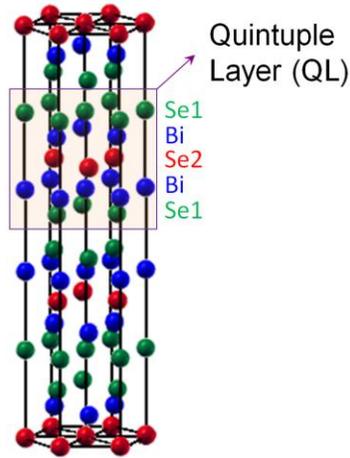

**Fig. 1**. Crystal structure of bulk $Bi_2Se_3$ hexagonal unit cell. A quintuple layer (QL), a basic building block of thin film structures, is indicated by the rectangle.

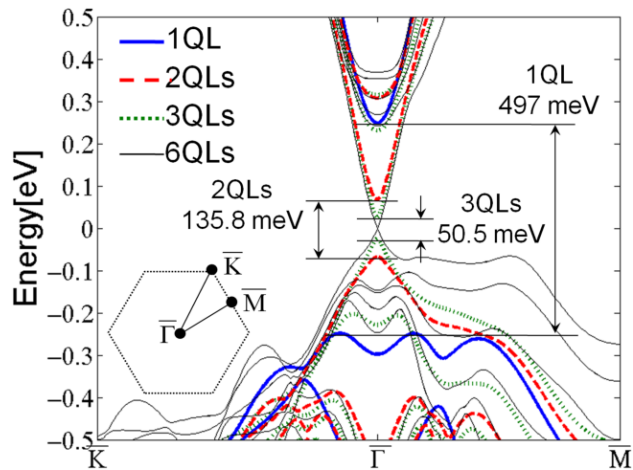

**Fig. 2**. Band structures of 1QL, 2QLs, 3QLs and 6QLs thin films obtained by DFT calculation along high symmetry directions in the hexagonal Brillouin Zone (BZ).

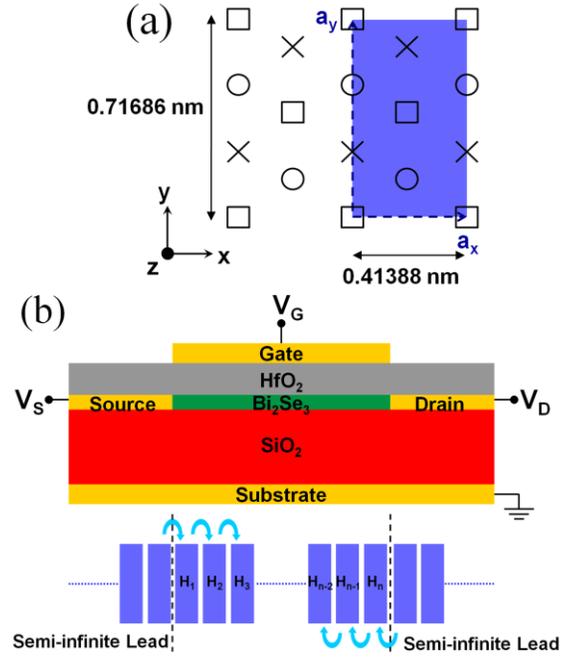

**Fig. 3**. (a) Top view of $Bi_2Se_3$ thin film. Three different symbols (○, □ and ×) are for atomic positions in different atomic layers stacked along the *z*-direction. A rectangular unit cell is denoted. (b) Device structure of $Bi_2Se_3$ MOSFET. The nominal device parameters are as follows: $Bi_2Se_3$ ($\kappa = 100$) thin film = 1QL (~0.7 nm), $HfO_2$ ($\kappa = 25$) gate oxide thickness = 2 nm, channel length = 20 and 50 nm, n-type doping density of source and drain = $1\times10^{13}$ $cm^{-2}$.

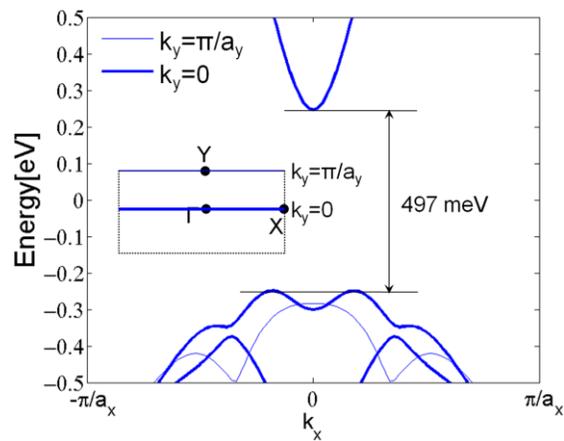

**Fig. 4**. Band structures of a 1QL thin film calculated from the TB Hamiltonian for two different transverse modes ($k_y=0$ and $\pi/a_y$) in the rectangular BZ.

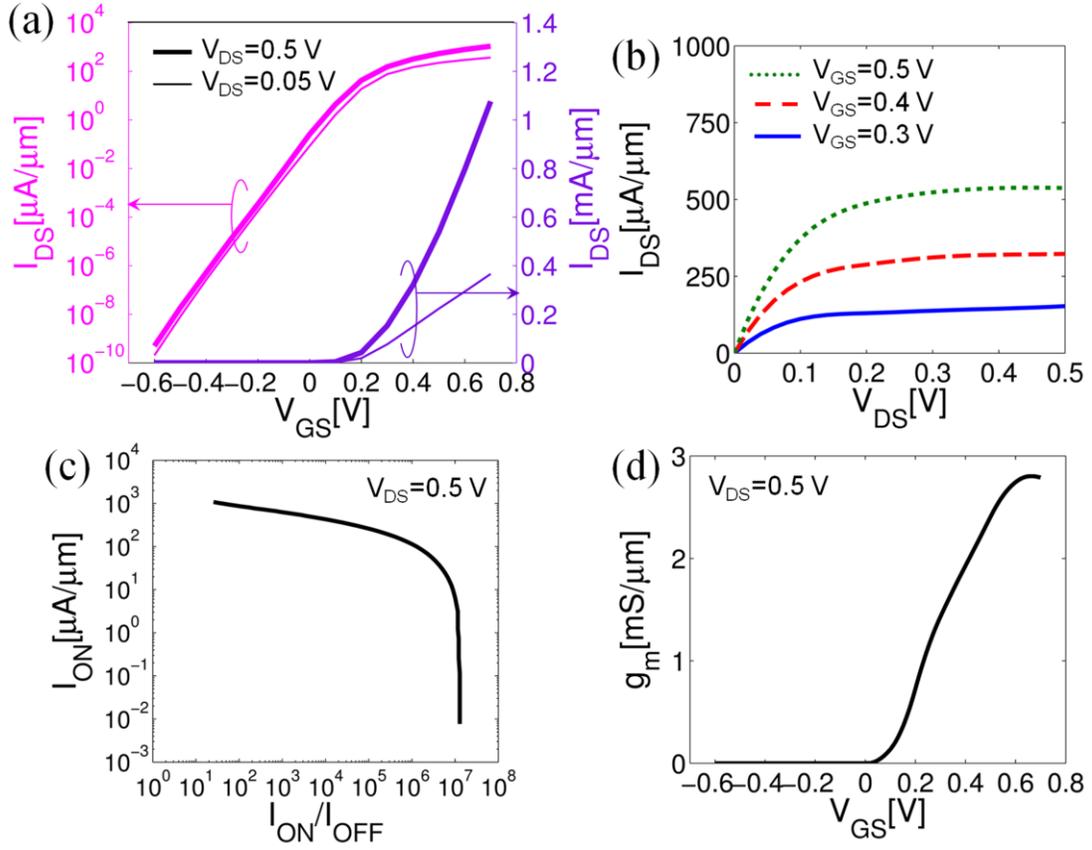

**Fig. 5**. Characteristics of 50 nm channel length 1QL Bi$_2$Se$_3$ MOSFET: (a) $I_{DS}$ vs. $V_{GS}$ curves at $V_{DS}$ = 0.05 V and $V_{DS}$ = 0.5 V on logarithmic (left axis) and linear scales (right axis), (b) $I_{DS}$ vs. $V_{DS}$ curves at $V_{GS}$ = 0.3, 0.4 and 0.5 V, (c) $I_{ON}$ vs. $I_{ON}/I_{OFF}$ at $V_{DS}$ = 0.5 V and (d) transconductance ($g_m = \partial I_{DS}/\partial V_{GS}$) vs. $V_{GS}$ at $V_{DS}$ = 0.5 V.

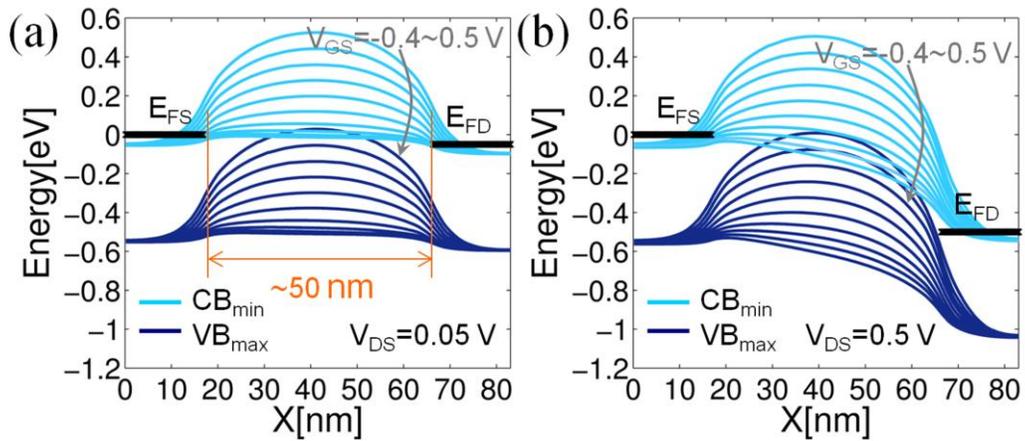

**Fig. 6**. CB and VB edges profiles along the 50 nm channel length $Bi_2Se_3$ MOSFET for $V_{GS}$ from −0.4 to 0.5 V in steps of 0.1 V (a) at $V_{DS}$ = 0.05 V and (b) $V_{DS}$ = 0.5 V.

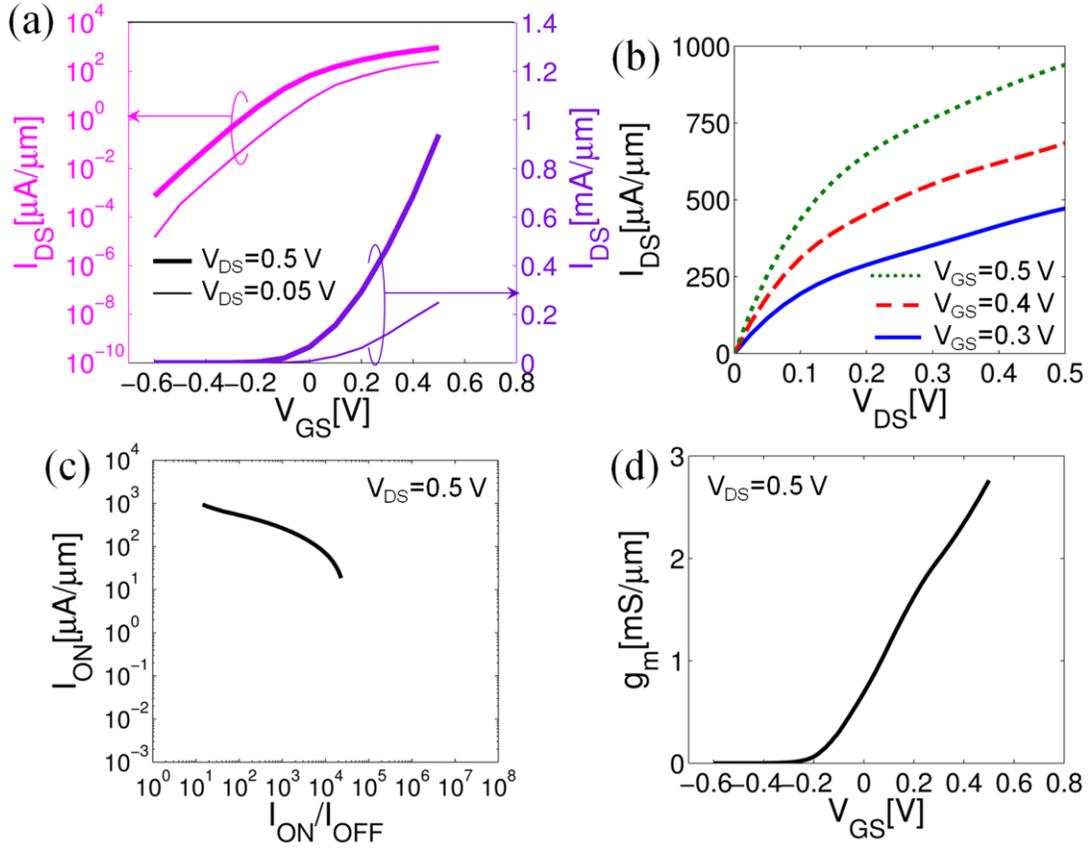

**Fig. 7**. Characteristics of 20 nm channel length 1QL $Bi_2Se_3$ MOSFET: (a) $I_{DS}$ vs. $V_{GS}$ curves at $V_{DS}$ = 0.05 V and $V_{DS}$ = 0.5 V on logarithmic (left axis) and linear scales (right axis), (b) $I_{DS}$ vs. $V_{DS}$ curves at $V_{GS}$ = 0.3, 0.4 and 0.5 V, (c) $I_{ON}$ vs. $I_{ON}/I_{OFF}$ at $V_{DS}$ = 0.5 V and (d) transconductance ($g_m = \partial I_{DS}/\partial V_{GS}$) vs. $V_{GS}$ at $V_{DS}$ = 0.5 V. All are plotted over the same ranges as for the 50 nm device of Figure 6 for better comparison.

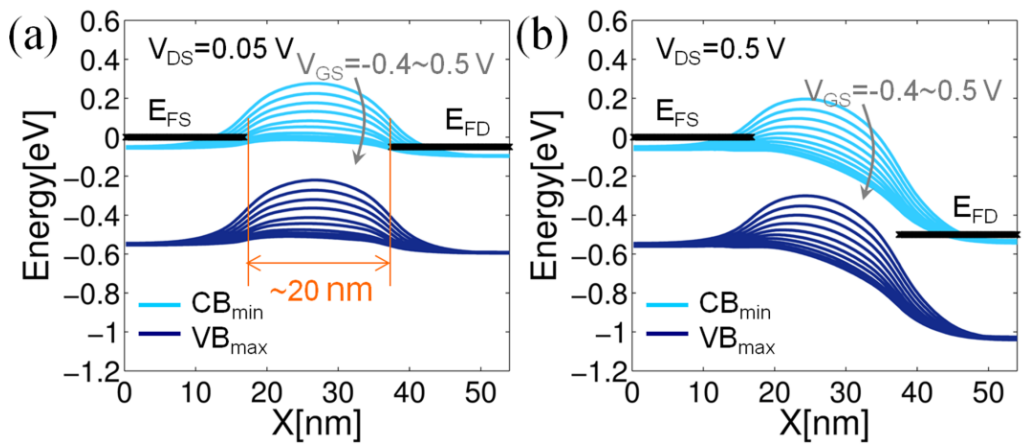

**Fig. 8**. CB and VB edges profiles along the 20 nm channel length $Bi_2Se_3$ MOSFET for $V_{GS}$ from −0.4 to 0.5 V in steps of 0.1 V (a) at $V_{DS}$ = 0.05 V and (b) $V_{DS}$ = 0.5 V.

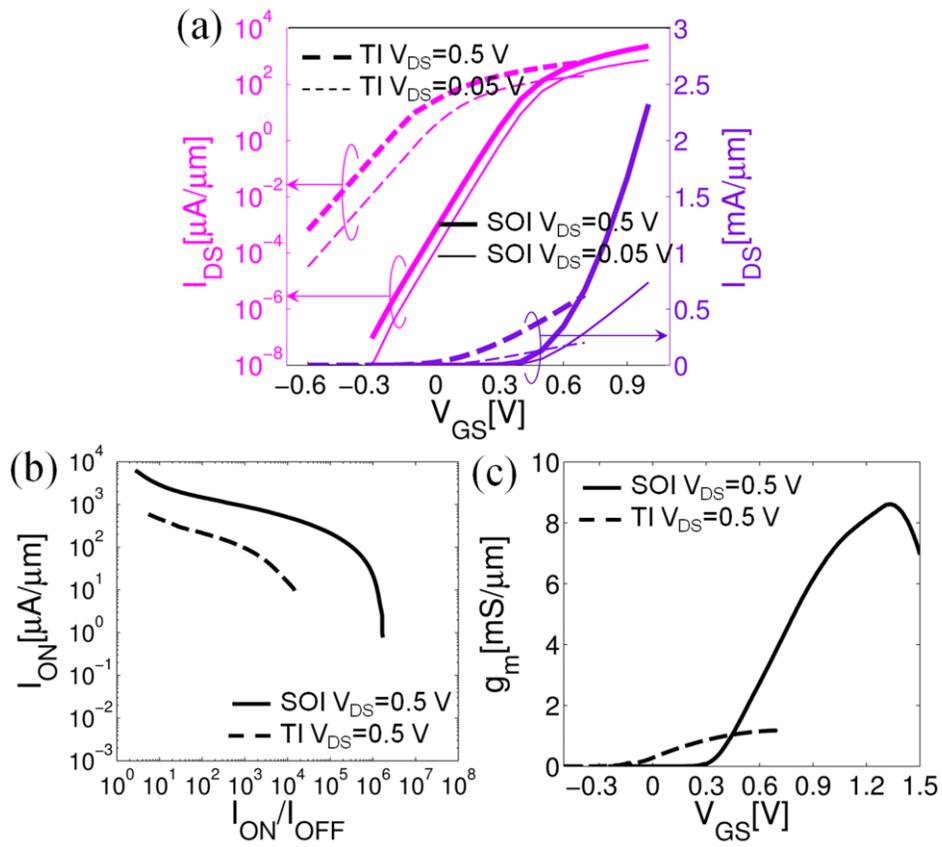

**Fig. 9**. Comparison of characteristics of 30 nm channel length SOI MOSFET and 1QL $Bi_2Se_3$ MOSFET (a) $I_{DS}$ vs. $V_{GS}$ curves at $V_{DS}$ = 0.05 V and $V_{DS}$ = 0.5 V on logarithmic (left axis) and linear scales (right axis), (b) $I_{ON}$ vs. $I_{ON}/I_{OFF}$ at $V_{DS}$ = 0.5 V and (c) transconductance ($g_m$ = $\partial I_{DS}/\partial V_{GS}$) vs. $V_{GS}$ at $V_{DS}$ = 0.5 V.